\documentclass[12pt]{article}
\usepackage{epsf}
\usepackage{a4wide}
\usepackage{supercite}

\begin{document}

\textit{\textbf{Title:}} \textbf{Identification of Characteristic Protein Folding Channels in a 
Coarse-Grained Hydrophobic-Polar Peptide Model}\\[5mm]
\textit{\textbf{Authors:}} Stefan Schnabel, Michael Bachmann, and Wolfhard Janke\\[5mm]
\textit{\textbf{Author affiliation:}} Institut f\"ur Theoretische Physik and Centre for Theoretical Sciences (NTZ),
Universit\"at Leipzig, Augustusplatz 10/11, D-04109 Leipzig, Germany\\[8mm]
\textit{\textbf{Corresponding author:}} Prof.\ Dr.\ Wolfhard Janke\\ 
\parbox[t]{2cm}{Address:}\hfill\parbox[t]{16cm}{Institut f\"ur Theoretische Physik,
Universit\"at Leipzig,\\ Augustusplatz 10/11, D-04109 Leipzig, Germany}\\[6mm]
\parbox[t]{2cm}{Phone:}\hfill\parbox[t]{16cm}{+49 341 9732725}\\
\parbox[t]{2cm}{Fax:}\hfill\parbox[t]{16cm}{+49 341 9732548}\\
\parbox[t]{2cm}{E-mail:}\hfill\parbox[t]{16cm}{Wolfhard.Janke@itp.uni-leipzig.de}\\[5mm]
\newpage
\section*{\it Abstract:}
Folding channels and free-energy landscapes of hydrophobic-polar heteropolymers
are discussed on the basis of a minimalistic off-lattice coarse-grained model.
We investigate how rearrangements of hydrophobic and polar monomers in a heteropolymer 
sequence lead to completely different folding behaviors. Studying three exemplified 
sequences with the same content of hydrophobic and polar residues, we can 
reproduce within this simple model two-state folding, folding through intermediates,
as well as metastability.  
\newpage
\section*{\it Text:}
The understanding of protein folding is one of the major challenges of modern interdisciplinary
science. Proteins are linear chains of amino acids connected by peptide bonds and typically 
consist of many hundreds of these acid residues. Except proline and glycine, all 
amino acids occurring in natural proteins possess a backbone with identical atomic composition,
while they differ strongly in their flexible side chains connected with the $C^\alpha$ carbon atom
in the backbone. Secondary structures, such as helices, sheets, and turns, are mainly formed
and stabilized by hydrogen bonds between backbone atoms, while side chains widely influence the 
three-dimensional conformation, i.e., the tertiary structure, of the protein. Roughly, side chains
can be characterized as polar or hydrophobic, depending on their chemical structure. Through
attractive polarisation effects with the aqueous environment, polar residues tend
to form hydrogen bonds with surrounding water molecules, whereas hydrophobic side chains
disturb the surrounding polar ``network'' and effectively attract each other. For this reason, 
hydrophobic monomers rather form a dense hydrophobic core in the interior of the protein, which is
surrounded by a shell of polar residues. It is widely believed that the hydrophobic effect
is the main driving force in the folding process of proteins, which in many cases happens 
spontaneously following the generation of the genetically coded amino acid sequence in the ribosome.

Computer simulations of protein folding are difficult, mainly for two reasons. Firstly, the folding 
process is so slow (microseconds to seconds) that molecular dynamics simulations of the whole
folding trajectory are currently still impossible employing realistic models. Secondly, one reason 
for the slow folding dynamics
is the assumed funnel-like free-energy landscape~\cite{onuchic0,clementi1,onuchic1} with ``hidden'' barriers. 
This means that in many cases 
probably no single order parameter or reaction coordinate exists that is suitable to describe
the folding channel(s) in the free-energy landscape. The problem is partly due to the enormous 
differences in the flexibilities of the degrees of freedom. While covalent bond lengths and bond
angles between covalent bonds are ``rigid'' (and fixed in several all-atom protein models),
dihedral torsional angles are much more flexible, although the flexibility is frequently hindered 
by strong torsional barriers.
In consequence, studies of folding kinetics and thermodynamics revealing the conformational transitions
accompanying the folding process are therefore also difficult using Monte Carlo methods.     
In particular, investigating effects of mutation and permutation of a wild-type amino acid sequence 
on folding channels is very time consuming when employing realistic protein models. For just this purpose,
simplified lattice and off-lattice coarse-grained models were designed, some of which only incorporate two types of 
amino acids: hydrophobic and polar residues~\cite{dill1,still1}. This will not allow the study of 
secondary structures, but it should be possible to focus on qualitative aspects that help to systemize the
understanding of tertiary heteropolymer folding, as, for example, hydrophobic-core 
formation~\cite{sorenson1,bj0a,bj1,bj0b,hsu1,liang1,baj2}. 

Another widely used class of models in studies of folding cooperativity are G\={o}
models, where the native conformation enters as input and the energy function is defined as the 
similarity of a conformation with the native fold~\cite{clementi2,li1,koga1,kaya1,kaya3,schonbrun1,head1}.  
This and similar simple structure-stabilizing models have been employed to gain a better insight into 
the folding kinetics of an important class of proteins -- the so-called two-state 
folders~\cite{fersht1,mayor1,ozkan1,kara1,fersht2,kaya2,weikl1,weikl2}. It should be stressed, however,
that the latter models are knowledge-based, i.e., that they are gauged for the sequence considered. 
This is not the case in the above-mentioned coarse-grained model used in the
present work. A big advantage of these more
physical models is that a variety of folding behaviors can be studied, because the kinetics is not
guided towards a certain structure. This enables the comparison of different, related
sequences and has particular implications for non-two-state folding and
metastability, the latter primarily concerning designed synthetic peptides or mutated biopolymers.

The main focus of this paper is on investigations of folding channels 
in dependence of a suitable ``order'' or system parameter that
allows for studies of cooperative, global changes~\cite{dill2,chan1} of the macrostates dominating 
the ensemble.
To this end, we employ the simplest coarse-grained off-lattice hydrophobic-polar heteropolymer 
model, the 
so-called AB model~\cite{still1}, and perform comparative studies of changes in folding channels 
by permuting AB sequences. 
\section*{Model and Definitions}
For the present study, we employ the AB model~\cite{still1} in its original form, with the exception
that the heteropolymer conformations can extend into three dimensions. In this model, only hydrophobic ($A$)
and hydrophilic or polar ($B$) monomers are distinguished. The reason is that these two classes
of amino acids are mainly responsible for the tertiary fold and the folding process is governed
by the hydrophobic forces and usually ends up in a conformation with a compact hydrophobic core 
surrounded by a polar shell.

In the following, we denote the spatial position of the $i$th monomer in a heteropolymer consisting 
of $N$ residues by ${\bf r}_i$, $i=1,\ldots,N$, and the vector connecting nonadjacent monomers $i$ and $j$ 
by ${\bf r}_{ij}$. For covalent bond vectors, we set $|{\bf r}_{i\,i+1}|=1$. The bending angle 
between monomers $k$, $k+1$, and $k+2$ is $\vartheta_k$ 
($0\le \vartheta_k\le \pi$) and $\sigma_i=A,B$ symbolizes the type of the monomer. In the
AB model~\cite{still1}, the energy of a conformation is given by
\begin{equation}
\label{eq:ab}
E = \frac{1}{4}\sum\limits_{k=1}^{N-2}(1-\cos \vartheta_k)+
 4\sum\limits_{i=1}^{N-2}\sum\limits_{j=i+2}^N\left(\frac{1}{r_{ij}^{12}}
-\frac{C(\sigma_i,\sigma_j)}{r_{ij}^6} \right),
\end{equation} 
where the first term is the bending energy and the sum runs over the $(N-2)$ bending angles of successive 
bond vectors. 
The second term partially competes with the bending barrier by a 
potential of Lennard-Jones type. It depends on the distance between monomers being non-adjacent 
along the chain and accounts for the influence of the AB sequence on the energy.
The long-range behavior is attractive for pairs of like monomers and repulsive for $AB$ pairs
of monomers:
\begin{equation}
\label{stillC}
C(\sigma_i,\sigma_j)=\left\{\begin{array}{cl}
+1, & \hspace{7mm} \sigma_i,\sigma_j=A,\\
+1/2, & \hspace{7mm} \sigma_i,\sigma_j=B,\\
-1/2,  & \hspace{7mm} \sigma_i\neq \sigma_j.\\
\end{array} \right.    
\end{equation}
Exploring this model by means of multicanonical Monte Carlo simulations~\cite{muca1,muca2} and a
spherical update mechanism described in Ref.~\cite{baj2}, 
we study in the following the folding properties of the three sequences listed in 
Table~\ref{tab:seqs}. This is a subset of deliberately designed sequences given in Ref.~\cite{irb1}.
All sequences have the same content of hydrophobic $A$ (14 each) and polar $B$ (6 each) residues -- only the 
order of the different types of monomers is exchanged. For each sequence, 10 independent
simulations (including estimation of the multicanonical weight factors) were performed and statistics
of $2\times 10^8$ conformations in each of the 10 production runs per sequence was accumulated.
A detailed analysis of thermodynamic and structural properties was recently performed in
a separate multicanonical Monte Carlo study~\cite{baj2,rem0}.  
\section*{Results and Discussion}
The folding process of proteins is necessarily accompanied by cooperative conformational changes.
Although not phase transitions in the strict sense, it should be expected that one or a few
parameters can be defined that enable the description of the structural ordering process. The 
number of degrees of freedom in most all-atom models is given by the dihedral torsional
backbone and side-chain angles. In coarse-grained C$^\alpha$ models as the AB model used in this study, 
the original dihedral angles are replaced by a set of virtual torsional and bond angles. In fact, the number
of degrees of freedom is not necessarily reduced in simplified off-lattice models. Therefore, 
the complexity of the space of degrees of freedom is comparable with more realistic models, and
it is also a challenge to identify a suitable order parameter for the folding in such minimalistic 
heteropolymer models. On the other hand, the computational simplicity of these models allows for a more systematic 
and efficient analysis of the heteropolymer folding process. 
In Fig.~\ref{figure1}, we
show the probability distributions $p_{\rm ang}(\Theta,\Phi)$ of all successive pairs of 
virtual bond angles $\Theta_i=\pi-\vartheta_i$ and torsion angles $\Phi_i$~\cite{rem1} for the exemplified 
AB sequence S3 at several temperatures. This plot can be considered as the AB analogue of the Ramachandran 
map for real proteins. Although this representation is not appropriate to describe the folding process, 
which will be rather complicated for this example as described later on, a few interesting features can
already be read off from this figure. At the temperature $T=0.3$, we observe two domains in this landscape, i.e., 
a structural pre-ordering has already taken place. The distribution is noticeably peaked for 
bond angles around 90$^\circ$ and torsion angles close to 0$^\circ$, i.e., almost perfectly planar 
{\em cis} conformations are favored in the ensemble as well as segments with bond angles between
60$^\circ$ and 70$^\circ$ for a broad distribution of torsion angles mainly between 40$^\circ$
and 100$^\circ$. The reason for the large width of the torsion-angle distribution in this
region is that the temperature is still to high for fine-structuring within the conformations. 
Explicit torsional barriers might stabilize these segments even at this temperature but are disregarded in the model.
Decreasing the temperature down to $T=0.1$, we see
that the landscape of this accumulated distribution of the degrees of freedom becomes very complex,
and the peaks are much sharper. In fact, close to $T\approx 0.1$, we observe a conformational
transition towards the formation of the ground states. Actually, the complexity of this landscape
can be understood better when considering the folding channels in the following, where we will see
that this heteropolymer exhibits metastability and therefore rather glassy behavior. A remarkable
aspect is the formation of the peaks in the bond-angle distribution at low temperatures close
to 60$^\circ$, 90$^\circ$, and 120$^\circ$, as these angles are typical base angles in
face-centered cubic crystals. As this concerns only segments of the conformations, the conformational
transition is actually not a crystallization. 
Concluding, distributions of degrees of freedom are not quite useful to describe the folding process.
For this reason it is necessary to define a suitable effective system parameter~\cite{du1,pande2}. A useful
choice will be discussed in the following.

In analogy to studies of the specific folding behavior in all-atom protein models~\cite{okamoto1,okamoto2}, 
we use here a generalized variant of the overlap order parameter as introduced in Ref.~\cite{baj2}.
The idea is to define a simple and computationally low-cost measure for the similarity of two conformations,
where the differences of the angular degrees of freedom are calculated~\cite{rem2}. In order to consider this 
parameter as kind of order parameter, it is useful to compare conformations ${\bf X}=({\bf r}_1,\ldots,{\bf r}_N)$ 
of the actual ensemble with
a suitable reference conformation ${\bf X}^{(0)}$, which is preferably chosen to be the global-energy minimum 
conformation. We define the overlap parameter as follows:
\begin{equation}
\label{eq:ov}
Q({\bf X},{\bf X}^{(0)})=1 - d({\bf X},{\bf X}^{(0)}). 
\end{equation}
With $N_b=N-2$ and $N_t=N-3$ being the respective numbers of 
bond angles $\Theta_i$ and torsional angles $\Phi_i$,
the angular deviation between the conformations is calculated according to
\begin{equation}
\label{eq:dparam}
d({\bf X},{\bf X}^{(0)})=\frac{1}{\pi(N_b+N_t)}\Bigg[
\sum\limits_{i=1}^{N_b}d_b\left(\Theta_i,\Theta^{(0)}_i\right)+
\max\left(\sum\limits_{i=1}^{N_t}d_t^-\left(\Phi_i,\Phi^{(0)}_i\right),
\sum\limits_{i=1}^{N_t}d_t^+\left(\Phi_i,\Phi^{(0)}_i\right)\right)\Bigg],
\end{equation}
where
\begin{eqnarray*}
\label{eq:dparamB}
d_b(\Theta_i,\Theta^{(0)}_i)&=&|\Theta_i-\Theta^{(0)}_i|,\nonumber\\
d_t^\pm(\Phi_i,\Phi^{(0)}_i)&=&{\rm min} \left(|\Phi_i\pm\Phi^{(0)}_i|,2\pi-|\Phi_i\pm\Phi^{(0)}_i| \right).
\end{eqnarray*}
Here we have taken into account that the AB model is invariant under the reflection symmetry
$\Phi_i\to-\Phi_i$. Thus, it is not useful to distinguish between reflection-symmetric
conformations and therefore only the larger overlap is considered.
Since $-\pi\le \Phi_i\le \pi$ and $0\le\Theta_i\le\pi$, the overlap is unity, if all angles 
of the conformations ${\bf X}$ and ${\bf X}^{(0)}$ coincide, else $0\le Q<1$. It should be noted that the average 
overlap of a random conformation with the corresponding reference state is for the sequences considered close to 
$\langle Q\rangle\approx 0.66$.
As a rule of thumb, it can be concluded that values $Q<0.8$ indicate weak or no significant similarity
of a given structure with the reference conformation. 

The global energy minimum conformations for the three sequences, which will be used as reference
states ${\bf X}^{(0)}$ in Eq.~(\ref{eq:ov}), are shown in Figs.~\ref{figure2}(a), (b), and~\ref{figure3}(a),
respectively. The conformation rendered in Fig.~\ref{figure3}(b) has a similar energy compared
with the one in Fig.~\ref{figure3}(a), but possesses a different geometry. This means that sequence
S3 exhibits a kind of metastable behavior at low temperatures.
The values of the lowest energies associated with the conformations in Figs.~\ref{figure2} and~\ref{figure3} 
are listed in Table~\ref{tab:seqs}. These minimum energies were
identified within the multicanonical simulations and are in perfect agreement with previous 
results~\cite{baj2} from energy-landscape paving (ELP) optimizations~\cite{hansmann1}. 

For the qualitative discussion of the folding behavior it is useful to consider the histogram
of energy $E$ and angular overlap $Q$ obtained from the multicanonical simulations,
\begin{equation}
\label{eq:hmuca}
H_{\rm muca}(E,Q) = \sum\limits_t\,\delta_{E,E({\bf X}_t)}\delta_{Q,Q({\bf X}_t,{\bf X}^{(0)})},
\end{equation} 
where the sum runs over all Monte Carlo sweeps $t$. 
In Figs.~\ref{figure4}(a)--(c), the multicanonical histograms $H_{\rm muca}(E,Q)$ 
are plotted for the three sequences listed in Table~\ref{tab:seqs}.
Ideally, multicanonical sampling yields
a constant energy distribution 
\begin{equation}
\label{eq:hflat}
h_{\rm muca}(E)= \int\limits_0^1dQ\,H_{\rm muca}(E,Q) = {\rm const.}
\end{equation}
An exemplified plot of the actual
multicanonical distribution $h_{\rm muca}(E)$ and the density of states $g(E)=h_{\rm muca}(E)/W(E)$ for the 
sequence S1 is shown in Fig.~\ref{figure5}, where $W(E)$ is the multicanonical weight factor.  
In consequence, the distribution $H_{\rm muca}(E,Q)$ can suitably be used to identify the folding channels, 
independently of temperature. This is more difficult with temperature-dependent canonical distributions 
$P(E,Q)$, which can, of course,
be obtained from $H_{\rm muca}(E,Q)$ by a simple reweighting procedure, $P(E,Q)\sim H_{\rm muca}(E,Q)g(E)\exp(-E/k_BT)$.
Nonetheless, it should be noted that, since there is a unique one-to-one correspondence between the average energy $\langle E\rangle$ and
temperature $T$, regions of changes in the monotonic behavior of $H_{\rm muca}(E,Q)$ can also be assigned a temperature, where
a conformational transition occurs.

Interpreting the ridges of the probability distributions in Fig.~\ref{figure4} as folding channels, it can 
clearly be seen that the heteropolymers exhibit noticeable differences in the folding behavior 
towards the native conformations (N). Considering natural proteins it would not be surprising that different
sequences of amino acids cause in many cases not only different native folds but also vary in their
folding behavior. Here we are considering, however, a highly minimalistic heteropolymer model and hitherto it
was not clear whether it would be possible to separate characteristic folding channels in this
simple model, but as Fig.~\ref{figure4} demonstrates, in fact, it is. For sequence S1, we 
identify in Fig.~\ref{figure4}(a) a typical 
two-state characteristics. Approaching from high energies (or high temperatures), the conformations in the ensemble D have an angular overlap $Q\approx 0.7$ with the lowest-energy reference state shown
in Fig.~\ref{figure2}(a), which means that there is no significant similarity with the reference
structure, i.e., the ensemble D consists mainly of unfolded peptides. For energies $E<-30$ a second branch opens. This channel (N) leads to the native conformation
(for which $Q=1$ and $E_{\rm min}\approx -33.8$). The constant-energy distribution, where the main  
and native-fold channels D and N coexist, exhibits two peaks noticeably separated by a well. Therefore, the
conformational transition between the channels looks first-order-like, which is typical for
two-state folding. The main channel D contains the ensemble of unfolded conformations,
whereas the native-fold channel N represents the folded states. 

The two-state behavior is confirmed by analysing the temperature dependence of the minima
in the free-energy landscape. The free energy as a function of the ``order'' parameter $Q$
at fixed temperature can be suitably defined as:
\begin{equation} 
\label{eq:freeE}
F(Q)=-k_BT\ln p(Q).
\end{equation}
In this expression,
\begin{equation}
\label{eq:pofq}
p(Q_0)= \int {\cal D}{\bf X}\, \delta(Q_0-Q({\bf X},{\bf X}^{(0)}))\,e^{-E({\bf X})/k_BT}
\end{equation}
is related to the probability of finding a conformation with a given value of $Q$ in the canonical ensemble
at temperature $T$. The formal integration runs over all possible conformations ${\bf X}$. In 
Fig.~\ref{figure6}(a), the free-energy landscape at various temperatures is shown for sequence S1. At
comparatively high temperatures ($T=0.4$), only the unfolded states ($Q\approx 0.71$)
in the main folding channel D dominate. Decreasing the temperature, the second (native-fold) channel N begins 
to form ($Q\approx 0.9$), but the global free-energy minimum is still associated with the main channel. 
Near $T\approx 0.1$, both free-energy minima have approximately the same value, the folding transition
occurs. The discontinuous character of this conformational transition is manifest by the existence
of the free-energy barrier between the two macrostates. For even smaller temperatures, the native-fold-like
conformations ($Q>0.95$) dominate and fold smoothly towards the $Q=1$ reference conformation, which is the lowest-energy
conformation found in the simulation.

A significantly different folding behavior is noticed for the heteropolymer with sequence S2. The 
corresponding multicanonical histogram is shown in Fig.~\ref{figure4}(b) and represents a folding
event through an intermediate macrostate. The main channel D bifurcates and a side channel I branches off
continuously. This branching is followed by the formation of a third channel N, which ends in the native fold. 
The unique $Q=1$-fold is plotted in Fig.~\ref{figure2}(b). The characteristics of folding-through-intermediates
is also confirmed by the free-energy landscapes as shown for this sequence in 
Fig.~\ref{figure6}(b) at different temperatures. Approaching from high energies, the ensemble of 
denatured conformations D ($Q\approx 0.76$) is dominant. Close to the transition temperature 
$T\approx 0.05$, the intermediary phase I is reached. The overlap of these intermediary conformations 
with the native fold is about $Q\approx 0.9$. Decreasing the temperature further below the native-folding
threshold close to $T=0.01$, the hydrophobic-core formation is finished and stable native-fold-like conformations with
$Q>0.97$ dominate (N).

The most extreme behavior of the three exemplified sequences is exhibited by the heteropolymer S3.
The main channel D does not decay in favor of a native-fold channel. In fact, we observe both, the formation
of {\em two} separate native-fold channels M$_1$ and M$_2$. Channel
M$_1$ advances towards the $Q=1$ fold as shown in Fig.~\ref{figure3}(a) and M$_2$ ends up 
in a completely different conformation with approximately the same energy, which is shown in
Fig.~\ref{figure3}(b). 
The spatial structures of these two conformations are noticeably different and their mutual overlap is
correspondingly very small, $Q\approx 0.746$. 
It should also be noted that the lowest-energy conformations in the main channel D
have only slightly larger energies than the two native folds. Thus, the folding of this heteropolymer
is accompanied by a very complex folding characteristics. In fact, this multiple-peak distribution
near minimum energies is a strong indication for metastability. A native fold in the natural sense
does not exist, the $Q=1$ conformation is only a reference state but the folding towards this
structure is not distinguished as it is in the folding characteristics of sequences S1 and S2.
This explains also, why the bond- and torsion-angle distribution in Fig.~\ref{figure1} possesses
so many spikes: it represents rather the ensemble of amorphous conformations than a distinct
footprint of a distinguished native fold. The amorphous folding behavior is also seen in the free-energy
landscapes in Fig.~\ref{figure6}(c). Above the folding transitions ($T=0.2$) the typical sequence-independent 
denatured conformations with $\langle Q\rangle \approx 0.77$ dominate (D). Then, in the annealing process, 
several channels are formed and coexist. The two most prominent channels (to which the lowest-energy
conformations belong that we found in the simulations) eventually lead for $T\approx 0.01$ to ensembles of macrostates
with $Q>0.97$ (M$_1$), which are similar to the reference conformation shown in Fig.~\ref{figure3}(a), and 
conformations with $Q<0.75$ (M$_2$). The lowest-energy conformation found in this regime is shown in 
Fig.~\ref{figure3}(b) and is structurally different but energetically degenerate compared with the 
reference conformation.
\section*{Summary}
The purpose of this paper is to show that even with a very simple coarse-grained
model, the AB model~\cite{still1}, complex sequence-dependent folding behavior 
of hydrophobic-polar heteropolymers can be investigated. This is very advantageous, since folding studies
of all-atom protein models with full force fields are very time consuming and a systematic,
comparative study of folding channels for a set of mutants would presently be still not
feasible with realistic efforts. Although
a one-to-one correspondence with real proteins cannot be claimed employing such
a minimalistic model, it is a useful tool for understanding general mechanisms
of tertiary heteropolymer folding. For this purpose, we have performed multicanonical simulations
for three exemplified hydrophobic-polar sequences. It should be noted that these
sequences are a subset of 20-mers studied within a different context~\cite{irb1} and, therefore, 
were not designed for the present study. Nonetheless, we found surprisingly complex
folding behaviors, which are, in fact, qualitatively comparable to known characteristics
of bioproteins and synthetic peptides. Beside the typical two-state folding behavior of sequence S1,
we also observed folding through an intermediate macrostate (sequence S2), as well as folding
into metastable conformations (sequence S3). 

Since the study of phase transitions in complex disordered systems such as, e.g., diluted ferromagnets, 
spin glasses, or structural glasses is successfully performed by means of simple
models, we expect that also the understanding of conformational transitions of
heteropolymers (which are intrinsically disordered by the nonhomogeneous sequence of hydrophobic and
hydrophilic monomers~\cite{wolynes1,pande1,pitard1}) can be advanced by studies of minimalistic models.

This work is partially supported by the DFG (German Science Foundation)  
under Grant contract No.\ JA 483/24-1. Some simulations were performed on the 
supercomputer JUMP of the John von Neumann Institute for Computing (NIC), Forschungszentrum
J\"ulich under grant No.\ hlz11. 
\newpage
\setlength{\parindent}{0em}

{\bf List of figure captions}\\[5mm]

\begin{figure*}
\centerline{\epsfxsize=17.6cm \epsfbox{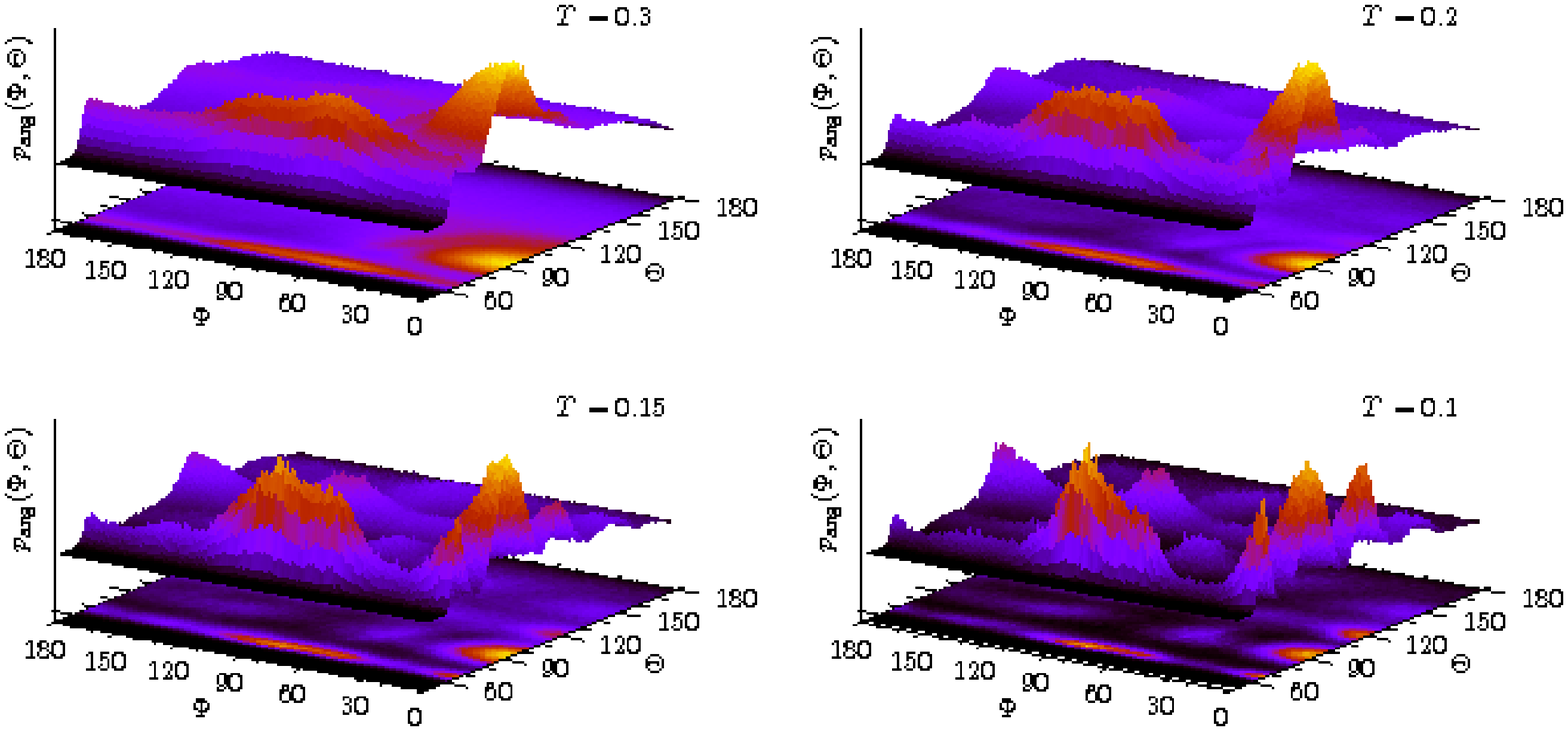}}
\caption{\label{figure1}(4 figures)}
\end{figure*}

{Fig.~\ref{figure1}:
(Color online) Bond and torsion angle distributions of sequence S3 at different
temperatures. The distributions of the torsion angles are reflection-symmetric and therefore only the 
positive intervals are shown. 
}

\begin{figure}
\centerline{\epsfxsize=8.8cm \epsfbox{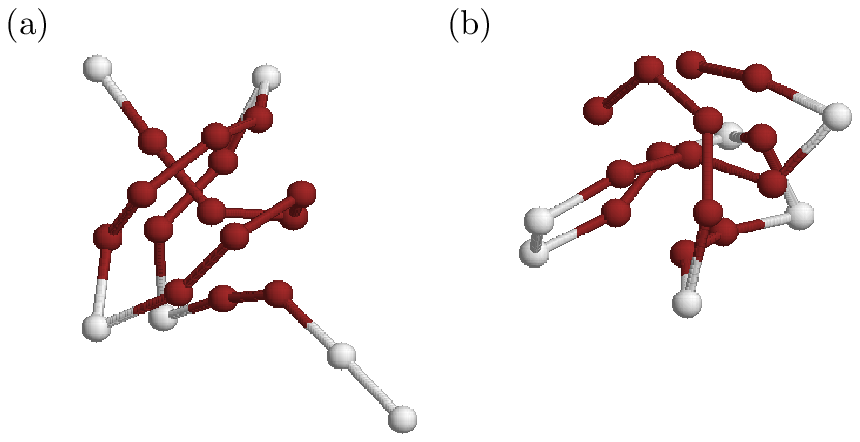}}
\caption{\label{figure2}(2 figures)}
\end{figure}

{Fig.~\ref{figure2}:
(Color online) Lowest-energy reference conformations ${\bf X}^{(0)}$ for sequences (a) S1 and (b) S2,
both residing in the respective native-fold channels (N).
}

\begin{figure}
\centerline{\epsfxsize=8.8cm \epsfbox{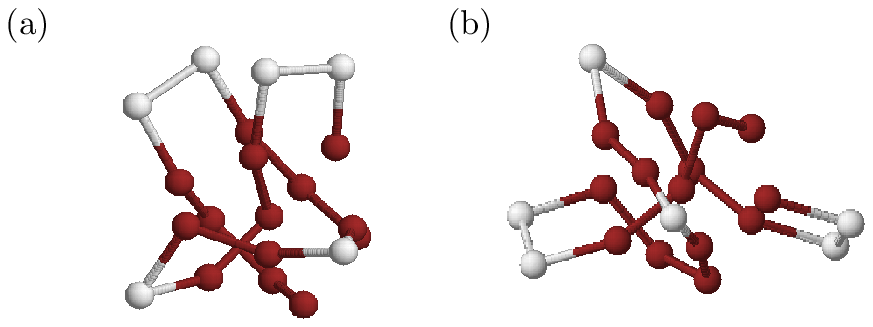}}
\caption{\label{figure3}(2 figures)}
\end{figure}

{Fig.~\ref{figure3}:
(Color online) Lowest-energy conformations for sequence S3, considered as (a) reference 
conformation ${\bf X}^{(0)}$ (M$_1$) and (b) alternative metastable conformation (M$_2$), 
whose angular overlap with ${\bf X}^{(0)}$ 
is $Q\approx 0.746$.
}

\begin{figure}
\centerline{\epsfxsize=8.8cm \epsfbox{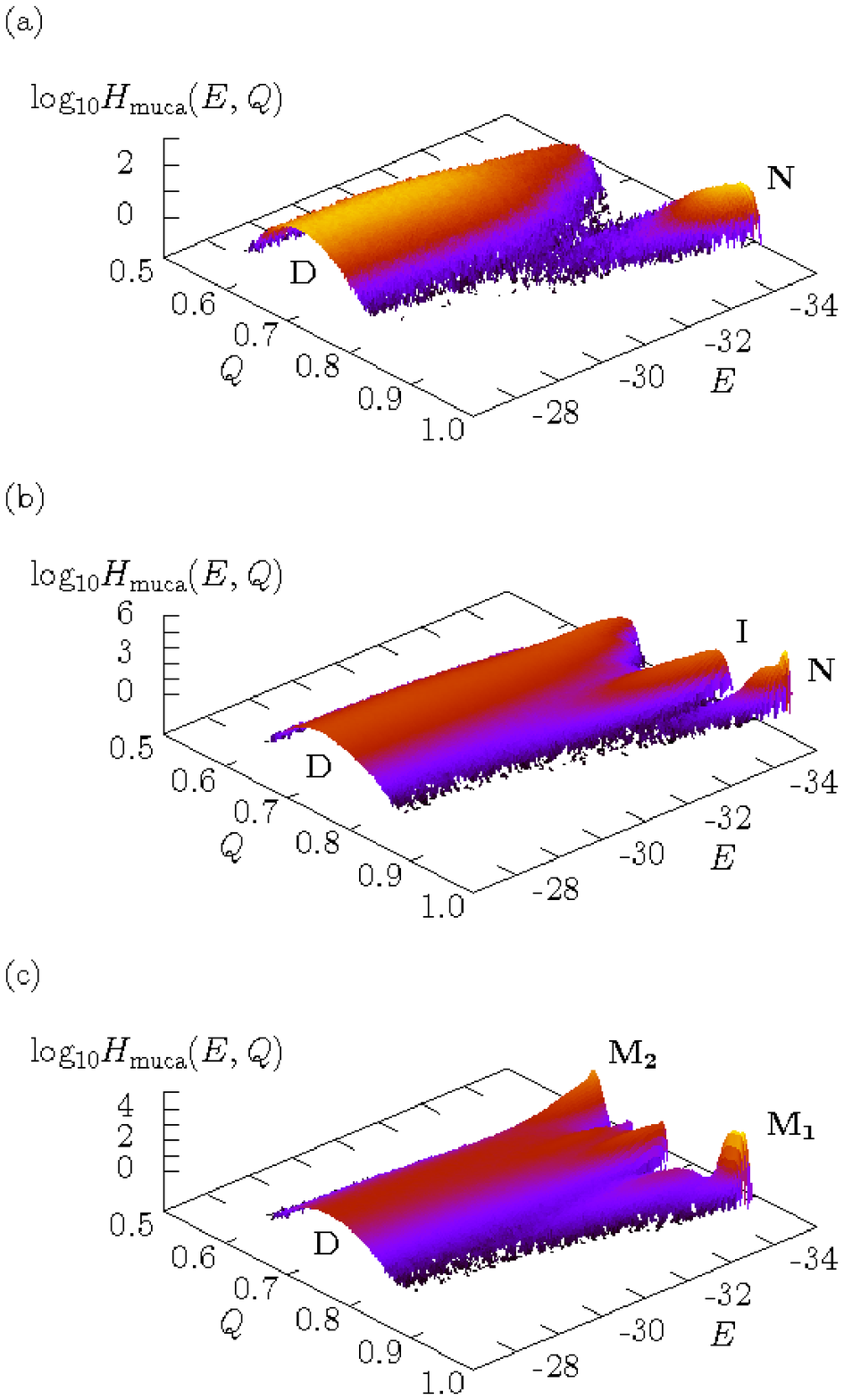}}
\caption{\label{figure4}(3 figures)}
\end{figure}

{Fig.~\ref{figure4}:
(Color online) Multicanonical histograms $H_{\rm muca}(E,Q)$ of energy $E$ and angular overlap parameter $Q$ 
for the three sequences (a) S1, (b) S2, and (c) S3. The different branches of these distributions are
channels the heteropolymer can follow in the folding process towards the native state. The native folds are located in the right 
corner for $Q=1$ and $E=E_{\rm min}$. Folding channels are labeled as D (denatured states), N (native
folds), I (intermediates), and M (metastable states).
}

\begin{figure}
\centerline{\epsfxsize=8.8cm \epsfbox{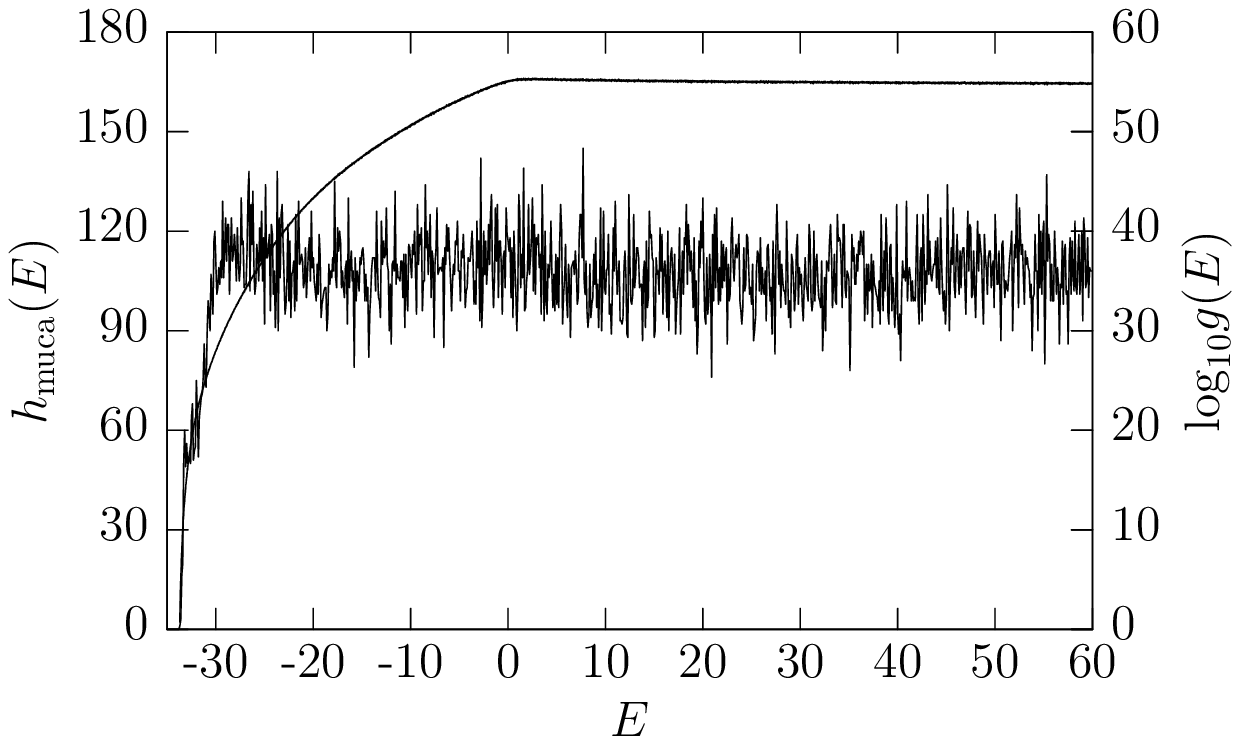}}
\caption{\label{figure5}(1 figure)}
\end{figure}

{Fig.~\ref{figure5}:
Multicanonical energy histogram $h_{\rm muca}(E)$ and density of states $g(E)$ 
for sequence S1.
}

\begin{figure}
\centerline{\epsfxsize=8.8cm \epsfbox{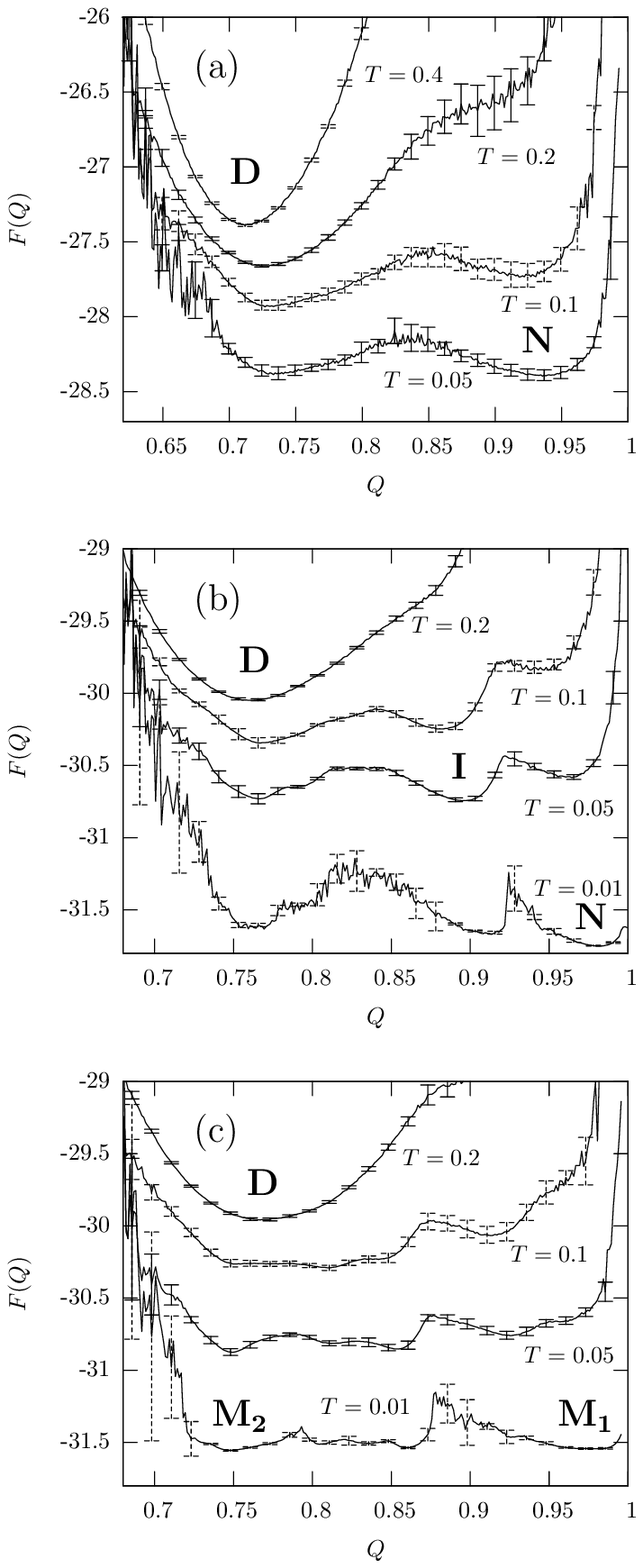}}
\caption{\label{figure6}(3 figures)}
\end{figure}

{Fig.~\ref{figure6}:
Free energy as a function of the overlap parameter at four different temperatures
for sequences (a) S1, (b) S2, and (c) S3. Note that the free energies are only determined up to a constant $F_0$
which was used to shift the curves for better discrimination. Labels of the free-energy minima refer to 
the folding channels in Fig.~\ref{figure4}.
}

\newpage
\clearpage
\newpage
\begin{table}[t]
\begin{center}
\caption{\label{tab:seqs} The three AB sequences with 20 monomers used in this paper
and the values of the associated global energy minima~\cite{rem0}. The two energies 
for S3 belong to qualitatively different conformations, which are considered in
the following as almost degenerate, metastable states (cf.\ Fig.~\ref{figure3}).}
\begin{tabular}{cp{3mm}cc}\hline\hline
label & & sequence & global energy minimum\\ \hline
S1 & & $BA_6BA_4BA_2BA_2B_2$ & $-33.8236$\\ 
S2 & & $A_4BA_2BABA_2B_2A_3BA_2$ & $-34.4892$\\ 
S3 & & $A_4B_2A_4BA_2BA_3B_2A$ & $-33.5838$, $-33.5116$\\ \hline \hline
\end{tabular}
\end{center}
\end{table}


\begin{thebibliography}{199}
\bibitem{rem0}
In the comparative multicanonical study of Ref.~\cite{baj2}, the sequence S1 was denoted 20.1, S2 had the label 20.4, 
and S3 was 20.3.
\bibitem{rem1}
For a polymer with $N$ monomers, there are $(N-2)$ bond and $(N-3)$ torsion angles. In order to form
successive pairs of these angles, we have left out the last bond angle (counted from the first monomer 
in the sequence) in Fig.~\ref{figure1}.
\bibitem{rem2}
Another, frequently used similarity measure is the root mean square deviation (rmsd) of two 
geometries. Since the rmsd calculation requires an optimization of the relative positions
of the main axes, this procedure is computationally more demanding than the calculation of the
overlap parameter used in this study.
\end{thebibliography}
\end{document}